\documentclass[12pt]{amsart}
\usepackage{geometry}                % See geometry.pdf to learn the layout options. There are lots.
\usepackage{graphicx}
\usepackage{amssymb}
\usepackage{epstopdf}
\usepackage{siunitx}
\title{Connecting particle clustering and rheology in attractive
particle networks}
\author{Sebastian Bindgen, Frank Bossler$^\ddag$, Jens Allard, Erin Koos*}
\date{\today \newline
KU Leuven, Chemical Engineering Department, Celestijnenlaan 200f, box 2424, 3001 Leuven, Belgium  E-mail: erin.koos@kuleuven.be
\newline
$^\ddag$ Present address: Zeppelinstraße 1c, 85375 Neufahrn, Germany
}                                           

\begin{document}

\maketitle

\begin{abstract}
The structural properties of suspensions and other multiphase systems are vital to overall processability, functionality and acceptance among consumers. Therefore, it is crucial to understand the intrinsic connection between the microstructure of a material and the resulting rheological properties. Here, we demonstrate how the transitions in the microstructural conformations can be quantified and correlated to rheological measurements. We find semi-local parameters from graph theory, the mathematical study of networks, to be useful in linking structure and rheology. Our results, using capillary suspensions as a model system, show that the use of the clustering coefficient, in combination with the coordination number, is able to capture not only the agglomeration of particles, but also measures the formation of groups. These phenomena are tightly connected to the rheological properties. The present sparse networks cannot be described by established techniques such as betweenness centrality.
\end{abstract}

\section{Introduction}
Multiphase systems, such as suspensions, provide a broad spectrum of applications and play an integral role in everyday life. Materials, like ketchup, wall paint and mud, are just a few examples. The physical properties of such suspensions are mostly influenced by their composition as well as interaction potentials and, hence, by the resulting microstructure. Phase and structural transitions are a key phenomenon regularly encountered in this field. Shear thinning, gelation, and agglomeration are phenomena that are often associated with microstructural changes in suspensions. The ability to tune and quantify this microstructure and, thereupon, relate it to the bulk properties is important for any application as well as rational and economic product design. A particular problem is the challenge of studying and controlling yielding due to the simultaneous dependence on local and global structures.

One hurdle in connecting structure and rheology originates from the presence of the many different length scales that are ubiquitous in the investigated systems. These length scales range from the typical length of interactions far below the particle's diameter, through the cluster scale spanning several particle diameters, to sample spanning scales.\cite{israelachvili2011intermolecular, cosgrove2005colloid} In the presence of attractive particle interactions the stability of larger clusters and connected structures on a more global scale is influenced by the type and strength of the attractive interaction and the ability of particle connections to transmit forces over larger length scales. Rheological phenomena, such as yielding, are influenced by the strength of the attractive interactions as well as the properties of the network such as the dimensionality. This poses a challenge to simultaneously quantify both scales and describe their relative importance.

Multiple tools have been developed to accomplish the task of describing the network structure. A considerable number of these methods rely on image analysis of confocal microscope images.\cite{prasad2007confocal, weeks2000three} One key property, often used in particle-based systems, is the coordination number, i.e. the number of contacts that a particle has with its neighbours. For dense systems such as granular matter, the coordination number can be related to mechanical rigidity and the jamming transition.\cite{papadopoulos2016evolution} In wet granular matter, it has been shown that the mechanical properties of a granular pile increase drastically with the addition of a small amount of liquid forming attractive bridges between neighbouring particles.\cite{Fournier2005, Scheel2008}  The coordination number was also the parameter of choice used by Dibble et al.\cite{dibble2006structure} amongst others to describe particle mobility in depletion gels with varying attraction strength as present in less clustered morphologies.\cite{weis2019structural}

Besides coordination number, several other network parameters have been investigated. In another depletion gel study, the rigidity of an attractive 3D particle network was characterised using the fractal dimension from image analysis.\cite{dinsmore2002direct} In this work, Dinsmore and Weitz suggested using pivot points and cluster sizes to further characterise different features of the structure. The coordination number was complemented with the void volume fraction to characterise structural heterogeneity in their samples.\cite{koumakis2015tuning} The role of network parameters, such as cycles or centrality measures, in granular matter has recently been extensively reviewed by Papadopoulos et al.\cite{papadopoulos2016evolution} Two other approaches, using the betweenness centrality in granular matter\cite{kollmer2019betweenness} and Cauchy-Born theory in depletion gels\cite{whitaker2019colloidal}, are discussed towards the end of our paper.

In short, the need for an easy, rigorous method to connect structure and rheology is only becoming more pressing with improved experimental and computational techniques. In computational experiments, dynamic properties are accessible using improved tracking mechanisms.\cite{besseling2009quantitative, fortini2012clustering} Experimentally, a method combining rheology and microscopy and including the measurement of local particle volume fractions was obtained using microrheology.\cite{chen2003rheological} Phase separation phenomena and the collapse of structures were also recently investigated.\cite{padmanabhan2018gravitational} The utilisation of knot theory showed the usefulness of mathematical concepts for nematic colloids.\cite{copar2015knot} Nevertheless, finding universal connections between the structure and mechanical properties of suspensions, especially embracing several length scales, remains challenging. 

The dimensionality of a network remains a rather arbitrary measure due to the requirement of spacial evenly distributed particle arrangements. Local differences such as dense clusters connected by thin bridges cannot be captured with such a technique and a clear prediction is not possible. The first attempt to successfully predict the modulus of such heterogenous networks showed that the in the region of heterogenous glasses the modulus is mostly dominated by intercluster contributions to the elastic response where multiple length scales become important.\cite{Zaccone2009}. Further work elaborated on stable rigid structures present in 2D samples that are dominated by triangles.\cite{zhang2019correlated}

The present study deals with tackling this challenge for the material class of capillary suspensions. Capillary suspensions are created by the addition of a secondary liquid to a two-phase suspension of particles in a bulk fluid, where the secondary and bulk liquids are immiscible.\cite{koos2011capillary} These materials are characterised by an overall structure consisting of a percolated network of particles  similar to colloidal gels, connected by attractive capillary interactions. While the capillary interactions are the same as in wet granular media, the smaller particle size decreases the {E\"{o}tv\"{o}s} number, the ratio of gravitational to capillary forces (also called the Bond number), from a value near unity for wet granular media to a value much, much smaller than unity for capillary suspensions.\cite{bossler2016structure} This difference in the relative strength of the capillary bridges over the particle weight can be readily observed in the typical solid loading for each system. Wet granular media tends to approach the volume fraction of random loose packing ($\phi_{\mathrm{solid}} \approx 0.5$) whereas capillary suspensions tend to have low to intermedia loading ($\phi_{\mathrm{solid}} \approx 0.1$--$0.4$).\cite{koos2014capillary}
This type of suspensions offers a wide domain of possible applications for the creation of smart materials.\cite{hauf2018structure, anupam2017thermally, bharti2015nanocapillarity, roh20193d}

Recent investigations on capillary suspensions have shown a broad range of options to tune the mechanical and rheological properties through changes in their composition,\cite{koos2012tuning} which are directly related to changes in their microstructure.\cite{bossler2016structure, dittmann2014micro, yang2017microstructure} Capillary suspensions show, in general, a gel-like behaviour with the storage modulus $G'$ larger than the loss modulus $G''$ and frequency independent behaviour inside of the linear viscoelastic region. Capillary suspensions exhibit a yield stress due to the strong attractive force transmitted by the secondary fluid, even in cases where the particles are not preferentially wetted by the secondary fluid.\cite{koos2014capillary} This yield stress shows a non-linear dependence on the amount of secondary fluid, indicating a complex relationship between the network microstructure and the overall network's mechanical response.\cite{bossler2018fractal} A confocal study has revealed several structural transitions in a silica model system when the wetting behaviour of both liquids on the particle surface is varied.\cite{bossler2016structure} The latter was accomplished by modifying the silica beads' chemical surface properties and, therefore, the three-phase contact angle $\theta$, which the secondary fluid forms towards the particles in a bulk liquid environment. In some cases, capillary suspensions can be characterised using scaling concepts that relate rheological data to the fractal dimension.\cite{bossler2018fractal} This work by Bossler et al. showed a network composed of aggregated flocs and different fractal models provided either the dimensionality of the floc\cite{wu2001model} or backbone\cite{shih1990scaling, piau1999shear}. These two fractal dimensions become larger with increasing particle size.\cite{bossler2018fractal} While the fractal dimension can be used to describe the connection between rheology and structure, this method can be problematic for heterogeneous network structures.\cite{bossler2018fractal} This problem becomes readily apparent in many capillary suspensions.

\begin{figure}[htbp]
\begin{center}
\includegraphics[width=0.5\textwidth]{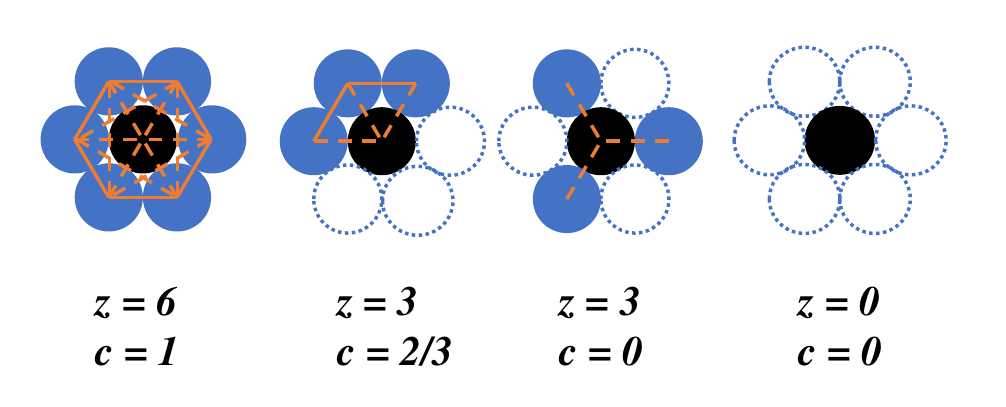}
\caption{Difference between the coordination number $z$, the number of bonds per particle (dashed lines), and the clustering coefficient $c$. Present particles are represented by filled circles. Dashed circles represent missing particles.}
\label{fig:clustering}
\end{center}
\end{figure}

While electron microscopy images of capillary-suspension-based ceramics microstructures showed structural transitions with different amounts of secondary liquid,\cite{domenech2017microstructure, dittmann2014micro, bossler2018fractal} no definitive measure exists for the connection between structure and rheology and thus transitions can be difficult to quantify. This is particularly true for the transition from a sparse network towards a more clustered or bicontinuous network for which measures such as the fractal dimension are ill-defined.\cite{koos2014capillary, yang2017microstructure, velankar2015non} Therefore, we need an approach that can work equally well for the sparse as well as dense networks. The number of neighbours, the coordination number $z$, which can be readily calculated from confocal images, provides some insights, but has a tenuous link to rheology. We therefore propose supplementing this measure with another measure from graph theory, namely the clustering coefficient, as shown in Figure~\ref{fig:clustering}. The local clustering coefficient is defined as 
\begin{equation}
c=\frac{2e}{z(z-1)},
\end{equation}
where the number of bonds between neighbours (dashed lines) is $z$ and the connections between these neighbours (solid lines) is $e$. An alternative definition of the clustering coefficient can be achieved by counting the number of triangles through a node (particle) in comparison with the coordination number of the corresponding node. This second definition will be used later on to quantify the stability of samples based on the Maxwell stability criterion.\cite{zhang2019correlated}
As illustrated in Figure~\ref{fig:clustering}, intermediate coordination numbers can either have low clustering ($c \rightarrow 0$) or high clustering ($c \rightarrow 1$) with clear implications of the mechanical response of the cluster. Although the upper limit of the clustering coefficient is $c=1$, as shown in the non-physical packed structure in Figure~\ref{fig:clustering}, real, e.g., biological networks, almost never reach this limit and $c>0.5$ is considered as high clustering.\cite{humphries2008network}

\section{Materials and Methods}
\subsection{Sample preparation}
We used spherical silica particles (Kromasil 100-7-SIL, Akzo Nobel) with mean particle radius of \SI{3.21}{\micro \meter} for the preparation of the capillary suspensions. The silica particles were covalently dyed with fluorescent rhodamine B isothiocyanate (Sigma-Aldrich) via a modified St\"{o}ber synthesis\cite{stober1968controlled, nozawa2005smart} and the particle surface was partially hydrophobized with two different concentrations of trimethylchlorosilane (Alfa Aesar) to achieve two different three-phase contact angles.\cite{fuji1998wettability} Details on the chemical modifications can be found in one of our recent publications.\cite{bossler2018fractal} Contact angles were determined from confocal microscopy images as $\theta = 87 \pm 8^\circ$ and $\theta = 115 \pm 8^\circ$.\cite{bossler2016structure} The bulk phase was a mixture of 86.8 wt \% 1,2-cyclohexane dicarboxylic acid diisononyl ester (Hexamoll DINCH, BASF) and 13.2 wt \% n-dodecane (Alfa Aesar). The secondary fluid was a mixture of $86.4$ wt \% glycerol (purity $> 99.5$ \%, Carl Roth) and $13.6$ wt \% ultrapure water. Minute amounts of the fluorescent dye PromoFluor-488 premium carboxylic acid (PromoKine) were added to the secondary fluid. The bulk and secondary fluid mixtures had the same refractive index as the dyed silica particles ($n = 1.455 \pm 0.001$). All capillary suspensions were prepared with the same effective solid volume fraction of $\phi_{\mathrm{solid}} = 20 \pm 4$ \%, which was directly determined from confocal images of the resulting samples. Varied amounts of secondary fluid were used. Since the particles are porous, the effective secondary fluid volume fraction $\phi_{\mathrm{sec}}$ was also identified from the confocal images. For sample preparation, the particles were first dispersed in the bulk fluid using ultrasonication, then the secondary fluid was added and ultrasound applied for another 30 s.\cite{bossler2018fractal} Total sample volumes were kept at 0.85 mL.

\subsection{Rheology} 
The rheological measurements were carried out with a stress-controlled Physica MCR 702 rheometer (Anton Paar) using a small-sized plate-plate geometry (8 mm plate diameter, 0.5 mm gap width). Strain sweep measurements were completed at a constant angular frequency of $\omega$ = 10 rad/s. Frequency sweep measurements were made in a frequency range of 100-0.1 rad/s at a previously determined strain amplitude within the linear viscoelastic (LVE) domain. In the observed range, $G’$ was frequency-independent in all cases. Thus, the plateau modulus $G^\prime_0$ was directly determined as the LVE value of $G’$ from the amplitude sweep, while the loss modulus $G''$ was determined at the critical amplitude $\hat\gamma_{\mathrm{crit}}$ denoting the end of the LVE region. All measurements were completed at 20 $^\circ$C. An example amplitude sweep for two secondary fluid volume fractions for the $\theta = 87 \pm 8^\circ$ sample is shown in the supplementary information Figure~S1.

\subsection{Confocal microscopy} 
3D confocal images (\SI{145x145x142 }{\micro \m}) were taken with a Leica TCS SP8 inverted confocal laser scanning microscope, equipped with two solid state lasers (488 nm and 552 nm). PromoFluor-488 premium carboxylic acid dye, in the capillary suspensions’ secondary liquid, is excited by the 488 nm laser and rhodamine B isothiocyanate, bonded to the particles, is excited by the 552 nm laser. The secondary phase dye emission was detected in a wavelength range of 495-520 nm and the particles at 650-700~nm. A glycerol immersion objective with a numerical aperture of 1.3 and 63$\times$ magnification was used for image acquisition. For every particle composition, eight image stacks were taken at different locations in the sample for a statistically appropriate sample size. The raw 3D images were used as input for the particle detection algorithm, which determined the individual particle locations and sizes, the effective volume fraction, and the coordination number distribution for each image.

\subsection{Particle detection, network reconstruction and graph theory approach}
The images were analysed using a self-written particle detection algorithm in IDL (version 8.7) using the rhodamine (particle) channel only. The particle positions were determined using a combination of Canny edge detection and Hough transform.\cite{canny1986computational} The Canny edge detection function for 2D images, available in IDL, was adapted for 3D confocal images. A Canny edge detector uses the derivative of a Gaussian function as edge detection kernel. This is equivalent to first performing a convolution on an image with a regular Gaussian, and subsequently taking the image gradient. In this way, the images were smoothed, thereby removing the detector noise from the images. A Gaussian kernel of $5 \times 5 \times 5$ pixels was applied with a sigma value of 1. The image gradient was taken in three dimensions to determine the gradient magnitude and direction. In the next step, the particle edges were trimmed down using non-maxima suppression, i.e. a pixel was retained if its gradient magnitude was a local maximum in the gradient direction. Hysteresis thresholding, specific to Canny edge detection, was not fully implemented. Only one threshold for the gradient magnitude, equal to 15 \% of the maximum gradient, was used to retain the strong (or true) edge pixels since edge linking was not the main objective and enough strong edge pixels remained to identify each particle.
After edge detection, a Hough transformed image was created from the strong edge pixels and the gradient direction. The availability of the gradient direction greatly simplifies the Hough transform since all edge pixel gradients point inwards to the particle center for spheres. Hence, the intermediate particle radius was the only parameter that was varied, in this case between \SI{1}{\micro \m} and \SI{6}{\micro \m} in steps of \SI{0.1}{\micro \m}. For each individual radius, the edge pixels casted a vote for a particle center in their gradient direction. The particle centres were determined as the pixels with the highest number of votes. These local maxima were detected using the 3D Crocker and Grier algorithm. The detected features were screened based on minimum number of votes, minimum brightness and overlap with other particles to remove misdetections. Finally, a least-squares fit for a sphere was performed on each detected feature (particle centres) and the edge pixels that voted for that feature (surface of the particle) using a Powell minimisation algorithm to determine the correct radius. Each sphere was characterised with four parameters: the x, y and z coordinates of the center and the particle radius. This algorithm offers an improved localisation and radius determination compared to the widely-used Crocker and Grier algorithm as required due to the large polydispersity and high volume fraction of the particles in the samples.\cite{crocker1996methods}

In order to analyse the network properties using graph theory, the 3D particle positions had to be reconstructed as a graph. This was done using the Python package Networkx.\cite{hagberg2008exploring} Each particle was mapped to a node. Node positions were assigned using an individual tag holding their coordinates. These nodes were sequentially connected by edges based on a distance criterion. We chose 6 pixels (\SI{0.85}{\micro \meter}) as threshold to detect bonds between individual particles and subsequently to set up edges between graph nodes. For the evaluation of number averages, we ignored particles closer than $2r$ to the edge of the pictures to ensure that only valid, fully evaluated particles are taken into account. All graphs present mean and standard deviations of the concatenated arrays over all samples. 

The result of the graph import procedure was an unweighted graph that was subsequently analysed.
We evaluated the coordination number $z$ and clustering coefficient $c$ for each node $n$.\cite{boccaletti2006complex, watts1998collective} The number averaged coordination number and clustering coefficient are denoted as $\bar{z} $ and $\bar{c}$, respectively.

\section{Results and Discussion}
\subsection{Microstructure} 
In the present study, two series of particles with differently treated surfaces, resulting in three-phase contact angles of $\theta = 87\pm8^\circ$ and $\theta = 115\pm8^\circ$, were prepared. The capillary suspensions with $\theta = 87\pm8^\circ$ are expected to show characteristics of the pendular state, i.e. a network stabilised by concave binary secondary fluid bridges that provide attractive capillary forces between the particles.\cite{koos2014capillary} Capillary suspensions with contact angles higher than $90^\circ$ are classified as being in the capillary state\cite{koos2011capillary} where the network structure is based on small aggregates made of a few particles that arrange around convexly shaped microdroplets of secondary fluid. Such a network gains its strength from an energetically favourability of forming such microclustered particle-droplet-arrangements.\cite{koos2012particle, fortini2012clustering} Hence, the samples with $\theta = 115\pm8^\circ$ are expected to represent the capillary state.  

\begin{figure}[htbp]
\begin{center}
\includegraphics[width=0.5\textwidth]{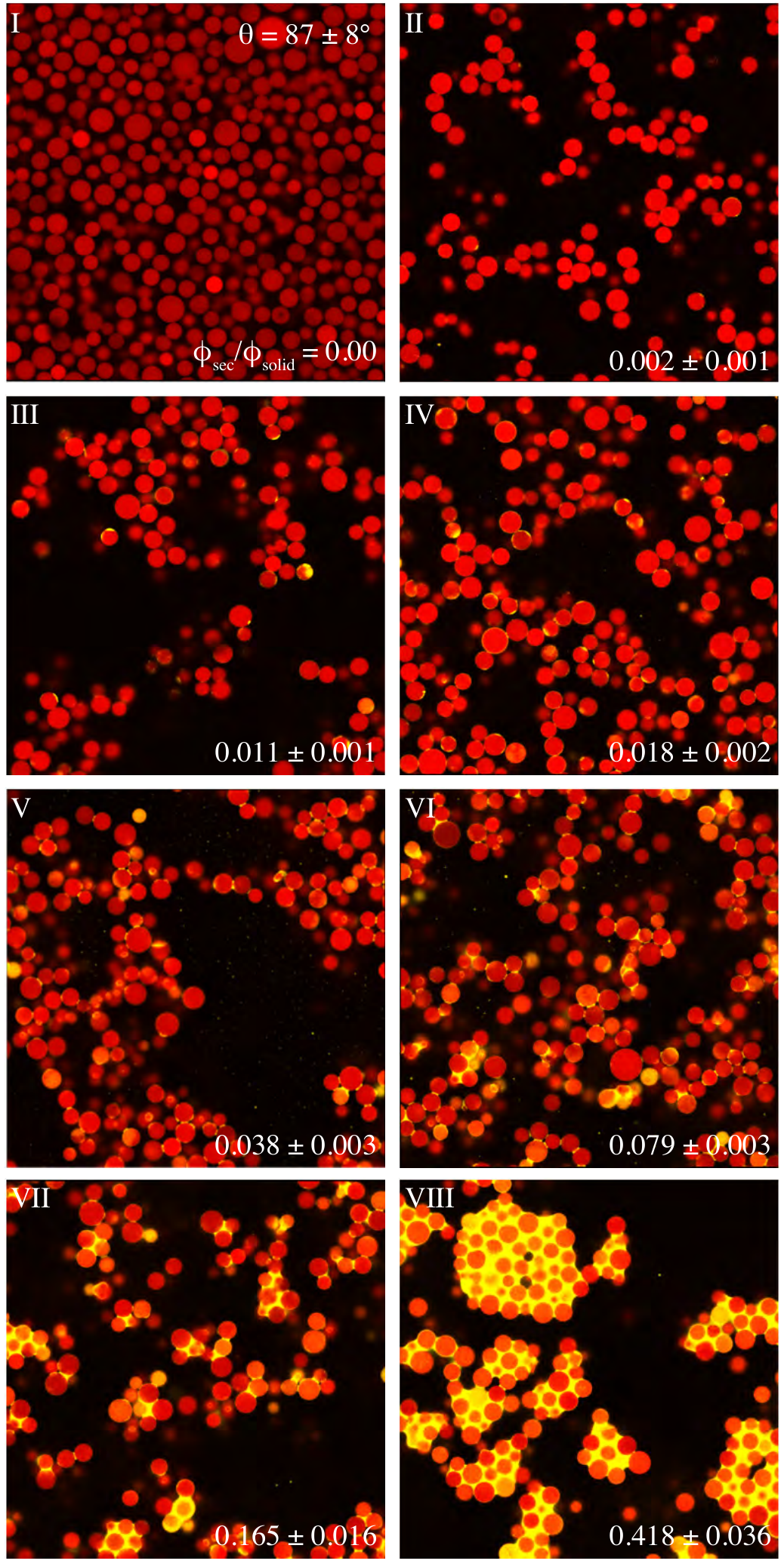}
\caption{Confocal 2D images of samples with contact angle $\theta = 87\pm8^\circ$ and varied amounts of secondary fluid. The particle detection channel is shown in red and the secondary fluid in yellow in these composite images. The size is \SI{145x145}{\micro \m}) in each image.}
\label{fig:Images_87}
\end{center}
\end{figure}

In order to better understand the transitions that occur with added secondary liquid, the sample microstructure is examined using confocal microscopy. Representative 2D slices of one of the eight 3D confocal images, taken for samples with varying $\phi_{\mathrm{sec}} / \phi_{\mathrm{solid}}$ for $\theta = 87\pm8^\circ$, are shown in Figure~\ref{fig:Images_87}.
The sample without added secondary fluid (Fig.~\ref{fig:Images_87}I) does not show a percolating particle network, but rather the structure of a loosely packed sedimented bed. This is easily visible by the much higher apparent solid volume fraction near the slide on the inverted microscope even though the same amount of particles was used during all sample preparations. This implies that any particle interactions are insufficient to overcome the weight of the particles. An addition of only $\phi_{\mathrm{sec}} / \phi_{\mathrm{solid}} = 0.002$ of secondary fluid leads to a completely different structure (Fig.~\ref{fig:Images_87}II). At such low secondary volume fractions $\phi_{\mathrm{sec}}$, individual particles are connected by binary contacts formed by bridges between asperities. The drops are small enough so that they are not visible in the image due to their size and low dye amount. However, the impact of the drops is clear since the particles form a network with small chains and flocs of particles. These flocs increase in size with increasing volumes of secondary fluid (Fig.~\ref{fig:Images_87}III and IV). The growth of clusters is linked first to the differing floc growth rate before percolation and second to droplet coalescence. Above a threshold volume, which depends on the contact angle $\theta$, the binary bridges on a particle trimer will touch and merge due to geometrical reasons.\cite{bossler2016structure, heidlebaugh2014aggregation} This is the point where aggregates of particles start to form and a funicular state results. For the $87 \pm 8^{\circ}$ contact angle in these images, this transition is predicted at a bridge volume of $V_{\mathrm{bridge}} / V_{\mathrm{particle}} = 0.03$. Assuming a coordination number of 2.5, this transition occurs at a $\phi_{\mathrm{sec}} / \phi_{\mathrm{solid}} = 0.04 \pm 0.02$. The merging of the bridges, i.e. a transition from a pendular state with only binary bridges to a funicular state with a clustered structure, is visible in Fig.~\ref{fig:Images_87}V and VI. Therefore, this transition takes place in the range of $0.038 \pm 0.003 \leq \phi_{\mathrm{sec}} / \phi_{\mathrm{solid}} \leq 0.079 \pm 0.003$, which correlates very well to the predicted value.
The aggregates grow in size and become more dense with further increase in $\phi_{\mathrm{sec}} / \phi_{\mathrm{solid}}$ until a bicontinuous structure of particles surrounded by secondary liquid, sometimes called a capillary cluster, is formed (Figure ~\ref{fig:Images_87}VIII).
While the network percolation in this confocal image is no longer visible directly, the large voids of bulk fluid between the particle agglomerates are proof of a stabilisation against sedimentation. This stabilisation is lost with a further increase to $\phi_{\mathrm{sec}} / \phi_{\mathrm{solid}} = 0.848 \pm 0.076$ when large, macroscopic agglomerates form (supplementary information Figure S2). 3D reconstructions of the networks are shown in the supplementary material Figure S3 to S5 where spheres are coloured according to their individual clustering coefficient and coordination number.

The $\theta = 115\pm8^\circ$ samples look similar to the ones with $\theta = 87\pm8^\circ$ at first glance, but they differ in some important ways. 
\begin{figure}[htbp]
\begin{center}
\includegraphics[width=0.5\textwidth]{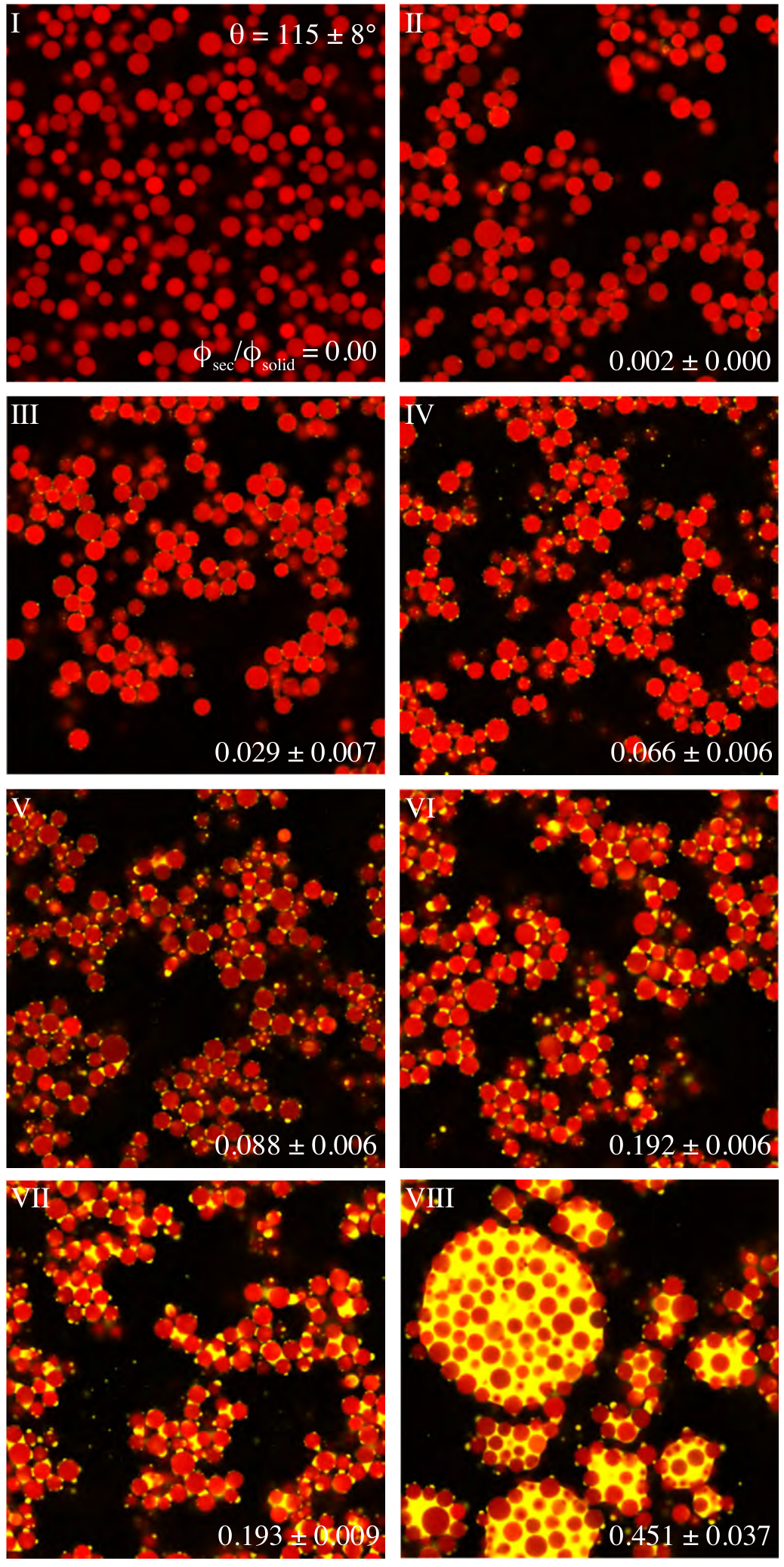}
\caption{Confocal 2D images of samples with contact angle $\theta = 115\pm8^\circ$ and varied amounts of secondary fluid.The particle detection channel is shown in red and the secondary fluid in yellow in these composite images. The size is \SI{145x145}{\micro \m} in each image.}
\label{fig:Images_115}
\end{center}
\end{figure}
First, the sample without added secondary fluid (Fig.~\ref{fig:Images_115}I) has a solid fraction larger than the average $\phi_{\mathrm{solid}} = 0.2$, but is less dense than for the lower contact angle. There may be a weak attractive interaction in this sample, meaning that the pure particle interactions in the bulk environment are different between the two contact angles, independent of any additional capillary force. This difference can be fully ascribed to the different surface modifications. 
While previous experiments have shown that the capillary state is less stable than the pendular state,\cite{bossler2016structure} this additional effect may help in supporting the network structure of the $\theta = 115 \pm 8^\circ$ samples here. 

With addition of increasing amounts of secondary fluid, the network formation in these samples is obvious (Fig.~\ref{fig:Images_115}II-IV). Network formation in the $115\pm8^\circ$ sample, however, appears to have a more clustered structure, even at small $\phi_{\mathrm{sec}} / \phi_{\mathrm{solid}}$ where individual drops are not visible, e.g. $\phi_{\mathrm{sec}} / \phi_{\mathrm{solid}} = 0.029 \pm 0.007$ (Figure~\ref{fig:Images_115}III). This means there is no transition from a linear, fractal-like network to a clustered structure, as was the case in the $\theta = 87\pm8^\circ$ sample. In the capillary state samples with $\theta > 90^\circ$, particles should aggregate around small droplets of secondary liquid. There are indeed capillary-state-like microclusters of particles around small droplets visible, as well as singly-connected droplets, which reside on particle surfaces without contributing to the network.\cite{bossler2016structure} Previously computational models of capillary state samples predict a transition from tetrahedral to larger particle number clusters with increasing amount of secondary fluid. The shear moduli and yield stress are expected to either reach a constant value or increase slightly in this region.\cite{koos2012particle}

Atypical for capillary state samples, however, is the visibility of unexpected binary bridges in Figure~\ref{fig:Images_115}V. It is also noteworthy that the particles form bicontinuous aggregates in the worse wetting secondary fluid instead of the better wetting bulk fluid at $\phi_{\mathrm{sec}} / \phi_{\mathrm{solid}} =0.451 \pm 0.037$ and in the phase separated clusters at $\phi_{\mathrm{sec}} / \phi_{\mathrm{solid}} =0.737 \pm 0.013$. These phenomenon may be related to the particle roughness causing a large contact angle hysteresis. 3D reconstructions of the networks are shown in the supplementary material Figure S6 to S8.

\subsection{Computational structure evaluations}
\subsubsection{Generic structures}
\begin{figure}[htbp]
\begin{center}
\includegraphics[width=0.4\textwidth]{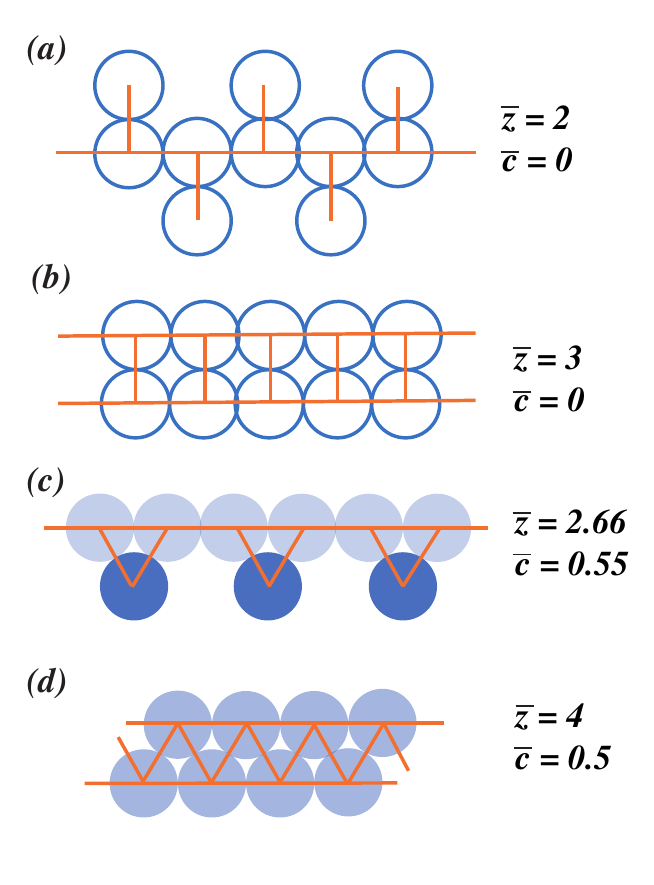}
\caption{Average coordination number $\bar{z}$ and clustering coefficient $\bar{c}$ for several structures commonly present in suspensions.The spheres are coloured from white to dark blue according to their individual clustering coefficient. These observations were made for 2D examples but }
\label{fig:examples}
\end{center}
\end{figure}

Several 2D structures are shown in Figure~\ref{fig:examples} to illustrate the dependence of the clustering coefficient and coordination number on the structure using several -- commonly encountered -- geometrical arrangements. These particle arrangements are representative segments of more extended structures generated by repeating these base units. 
Earlier results deducted from 2D samples and strictly radial force fields indicate that the stability of such structures can be described by the amount of triangles (Maxwell rigidity)\cite{zhang2019correlated}. We argue that the triangular structures detected by the clustering coefficient should correspond to rigid structures in 3D and, as such, allow a prediction of the overall modulus of the sample.

The example segments are shown in Figure~\ref{fig:examples}, where we display
a string of particles with small side branches (Figure~\ref{fig:examples}a), a square backbone arrangement (Figure~\ref{fig:examples}b), a string of particles with more tightly incorporated side branches (Figure~\ref{fig:examples}c), and particles with a triangular backbone (Figure~\ref{fig:examples}d). The square grid arrangements, illustrated in Figure~\ref{fig:examples}a and ~\ref{fig:examples}b both have $\bar{c} = 0$ despite the difference in the number of side branches and corresponding $\bar{z}$. In both cases, there are no triangles formed by bonds between neighbouring particles. These bonds between neighbouring particles, and hence a nonzero clustering coefficient, can be seen in examples Figures ~\ref{fig:examples}c and ~\ref{fig:examples}d. In example ~\ref{fig:examples}c, a structure is introduced with a main path connecting several, more tightly incorporated side beads. Whereas the main path has rather lower clustering, the side beads are fully tight to the graph forming cliques due to the formation of all possible triangles. Example ~\ref{fig:examples}d shows a triangular backbone of particles. Here, an average clustering coefficient of $\bar{c} =0.5$ is visible. This threshold is crucial for the later considerations in the histogram, because it allows us to distinguish thicker structures in Figure~\ref{fig:examples}d from thin structures as formed in Figure~\ref{fig:examples}c.

\subsubsection{Sample characterisation}
\begin{figure}[htbp]
\begin{center}
\includegraphics[width=0.5\textwidth]{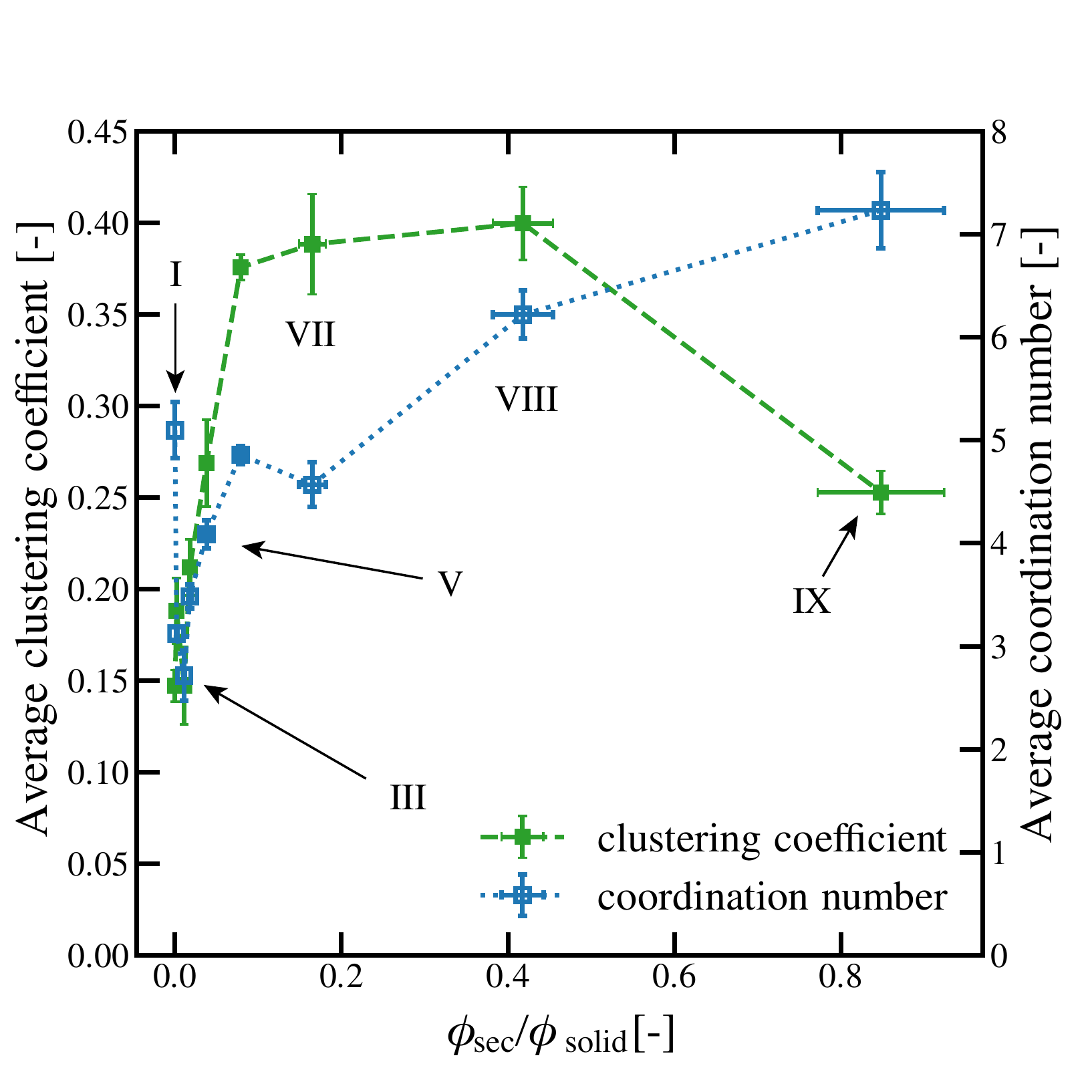}
\caption{Average coordination number and clustering coefficient for $\theta = 87\pm8^\circ$. The roman numerals correspond to the images shown in Figure~\ref{fig:Images_87}}
\label{fig:averages_pend}
\end{center}
\end{figure}

The coordination number and the clustering coefficient are used to quantify the previously described characteristics of the confocal images. The averaged values are shown in Figure~\ref{fig:averages_pend}. These values represent the average determined from all eight of the 3D images (or rather, the average of the 8 graphs). With the addition of secondary liquid, $\bar{z}$ decreases, reaching a minimum at point III, whereas $\bar{c}$ increases slightly. With more secondary fluid, both values then increase, with $\bar{c}$ reaching a plateau at point VI. The coordination number still increases in this range, but it has reached an inflection point. Finally, at point IX, the average clustering coefficient decreases with the transition to a granular aggregate.
A more detailed view on the first five points in this graph is shown in the supplementary material Figure S9.

\begin{figure*}[htbp]
\begin{center}
\includegraphics[width=\textwidth]{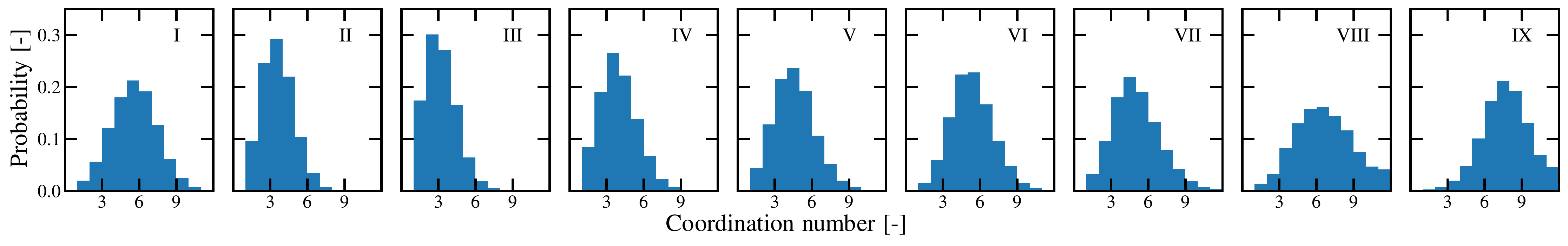}
\includegraphics[width=\textwidth]{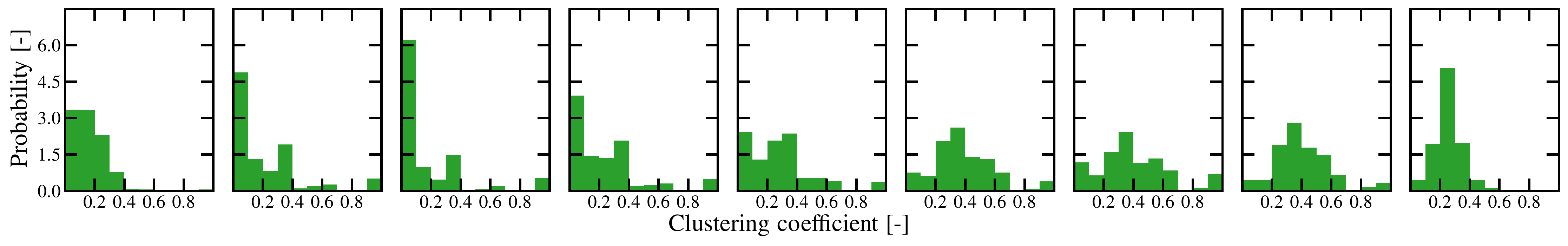}
\caption{Histograms of the coordination number and clustering coefficient for all samples with a contact angle of $\theta = 87\pm8^\circ$ for several amounts of secondary fluid. }
\label{fig:histograms_87}
\end{center}
\end{figure*}

While the average values give some indication for the changes in the network structure, key transitions can be hidden by subtle changes in the average values. The histograms of all samples are therefore shown in Figure~\ref{fig:histograms_87}.
These histograms will be used to show key transitions in the microstructure and finally to tie these changes to the resulting rheological properties. For all histograms, the clustering coefficient is shown in green and the coordination number in blue. All shown histograms are normalised so that the integral value over all bins is 1. A comparison between low ($ c \leq 0.2$), medium ($0.2 \leq c < 0.4$), high ($ 0.4 \leq c < 0.6 $) as well as the remaining ($ 0.6 \leq c$) clustering coefficient is shown in supplementary Figure S10.

We will first consider the sample that corresponds to the pendular state samples ($\theta = 87 \pm 8^\circ$). A broad distribution of coordination numbers can be seen for the sample without added secondary liquid (Figure~\ref{fig:histograms_87}I) with a mean value of $\bar{z}=5.1$. The mean value of the clustering coefficient is $\bar{c}=0.15$. A lot of particles with a clustering coefficient around the theoretical minimum of $c=0$ can be observed, indicating particles without tight clustering. These values are characteristic for a loosely packed sedimented bed. That means there is a random arrangement of particles present where polydispersity of the particles can lead to a lot of isolated particles or particles with only one neighbour. This low clustering can indicate a metastable structure.\cite{heylen2006metastable} Indeed, the structure of this sample changed slightly during the measurements, either due to the small movements of the objective or very slow sedimentation. 

With the addition of secondary fluid to $\phi_{\mathrm{sec}} / \phi_{\mathrm{solid}} = 0.002 \pm 0.001$ (point II), we can see the structural arrangements due to the filling of the particles asperities in the histograms. Particles are connected to each other via binary bridges leading to a lower average contact number of $\bar{z}=3.1$ while the average clustering coefficient rises to $\bar{c}=0.19$. This rise is caused by two effects. First, a number of particles with a high clustering coefficient close to $c=1$ appears. These highly clustered particles are arrangements of trimers or other small groups of particles, often located on the outside of flocs or strands. Second, a new peak emerges at an intermediate value of the clustering coefficient around $c=0.3$, which is consistent with particles in a partially reinforced backbone chain as illustrated by Figure~\ref{fig:examples}b. However, one must not ignore the rise in the amount of particles having low or zero clustering. We conclude from the corresponding histogram of the coordination number that there is a change from the random sediment arrangement into an open network with many short, linear branches of two or three, as indicated by the increase of particles with coordination number $z=1$ to $z=3$.

The same trend continues at $\phi_{\mathrm{sec}} / \phi_{\mathrm{solid}} = 0.011 \pm 0.001$ (Figure~\ref{fig:Images_87}III), where a fractal-like gel structure is visible. The average coordination number once more lowers to about $\bar{z}=2.7$. Furthermore, the amount of particles with a coordination number of $z=2$ reaches its global maximum compared with the other histograms. The distribution of the clustering coefficient, which has an average of $\bar{c}=0.15$, shows the same features as above but is even more pronounced. As the number of particles with low clustering coefficient rises, more binary contacts are present. This rise is associated with a decrease in the number of particles with intermediate clustering, indicating a change towards the linear branched structure of Figure~\ref{fig:examples}a. The amount of particles at the maximum clustering coefficient shows no change (see also Figure S2). Such a transition would be consistent with the formation of a weak gel structure dominated by binary contacts (Figure~\ref{fig:examples}a) and some tightly arranged clusters formed by trimers or other low number particle groups (Figure~\ref{fig:examples}b).

As the structure changes towards a funicular, or clustered, state (Figure~\ref{fig:Images_87}IV), the coordination number increases to a value of $\bar{z}=3.5$. The average clustering coefficient also increases to $\bar{c}=0.21$, but the change in the average, shown in Figure~\ref{fig:averages_pend}, is quite modest. The histogram shows a clearer loss in the number of particles with very low clustering and an increase in the intermediate clustering. We believe that this is caused by the addition of particles to already existing clusters, leading to their increase in size and not by the formation of entirely new groups of particles. This funicular state is akin to the sketch in Figure~\ref{fig:examples}b with the size of the triangular (or, in 3D, tetrahedral) clusters growing in size.

With the next three additions of secondary liquid, corresponding to the growth of funicular clusters and the transition to a fully bicontinuous structure, the average coordination number has values of $\bar{z}=4.1$, $\bar{z}=4.9$ and $\bar{z}=4.6$, respectively.  The clustering coefficient approaches a plateau with average values of $\bar{c}=0.26$, $\bar{c}=0.38$ and $\bar{c}=0.39$. These are values that are typical of small-world phenomena in graph theory. Small world graphs have nodes that are not direct neighbours (low $z$ or degree) but are only separated from each other by a few hops (a few particles). Put another way, these graphs are composed of tight cliques with few connections between cliques.  These maximal cliques are defined as groups of nodes (particles) where every two nodes are adjacent (forming a triangle) and the number of triangles would not be increased by including more nodes. Thus, the cliques are tight flocs of particles with few inter-floc connections.
A look at the confocal images reveals the presence of particle groups that become more and more isolated, indeed forming the small interconnected worlds. The confocal picture with $\phi_{\mathrm{sec}} / \phi_{\mathrm{solid}} = 0.418 \pm 0.036$ (Figure~\ref{fig:Images_87}VIII) especially illustrates that state.

With the final addition of secondary fluid, the structure phase separates. This is indicated by the clustering coefficient decreasing to an average value of $\bar{c} = 0.25$, towards the initial sedimented bed state. However, the structure seems to be denser than the initial sediment, as shown by the increased clustering coefficient and coordination number ($\bar{z}=7.2$). This can be explained by the fact that the secondary liquid is solely present between the sedimented particles, pulling them together more tightly. This means that the arrangement of particles is not as loosely arranged as in the first state without secondary liquid added. This structure should be more resilient to external deformation.

\begin{figure}[htbp]
\begin{center}
\includegraphics[width=0.5\textwidth]{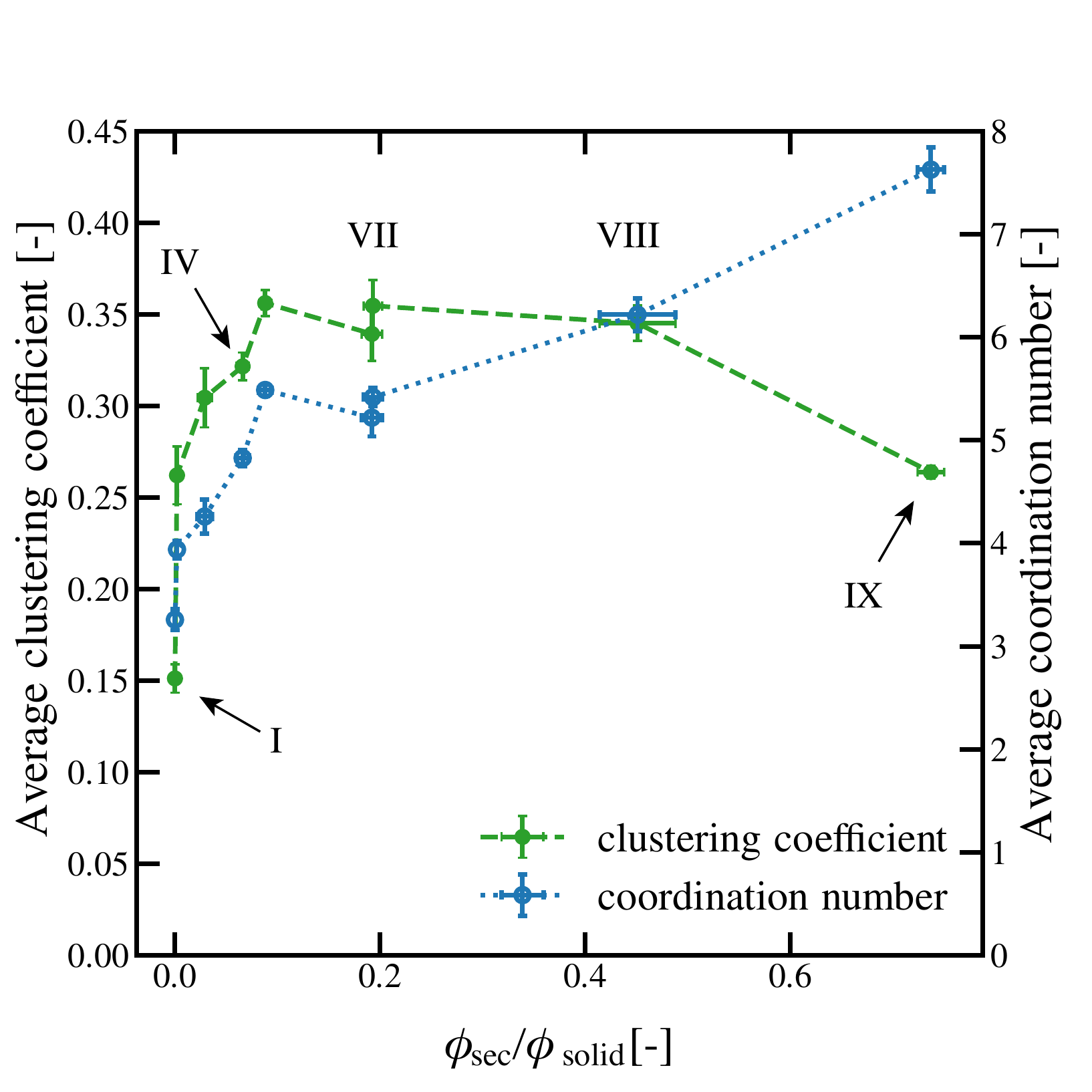}
\caption{Average coordination number and clustering coefficient for $\theta = 115\pm8^\circ$. The roman numeral correspond to the images shown in Figure~\ref{fig:Images_115}}
\label{fig:averages_cap}
\end{center}
\end{figure}

\begin{figure*}[htbp]
\begin{center}
\includegraphics[width=\textwidth]{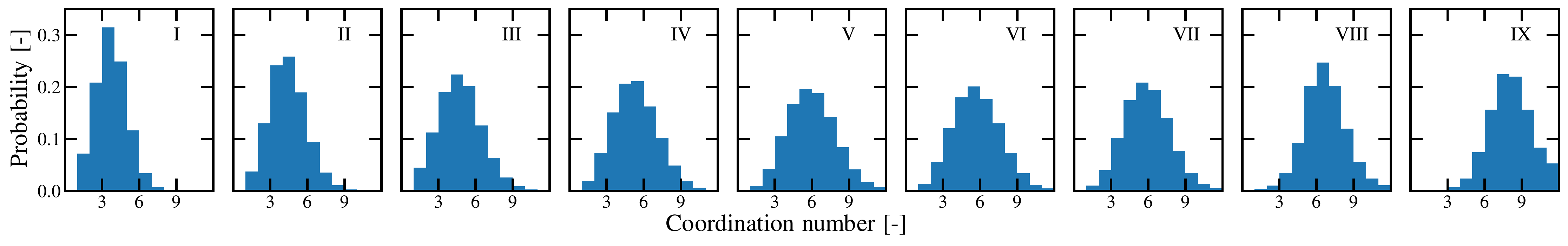}
\includegraphics[width=\textwidth]{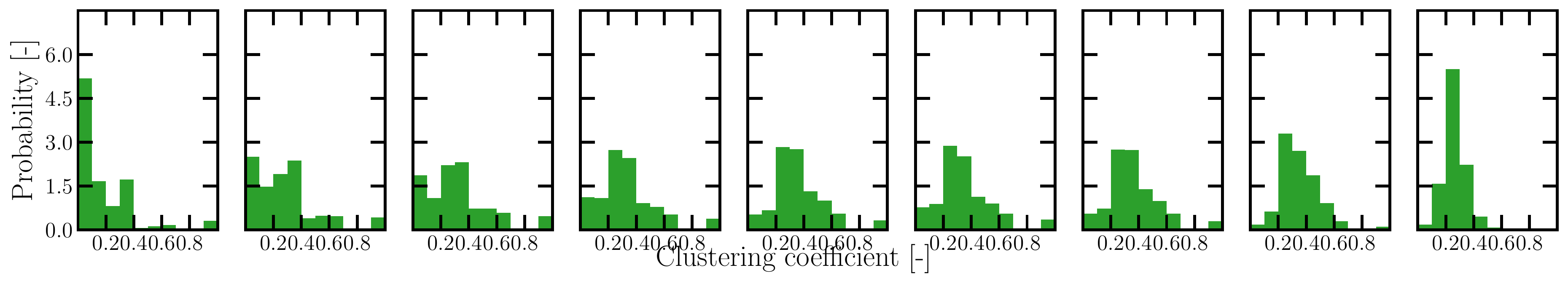}
\caption{Histograms of the coordination number and clustering coefficient for all samples with a contact angle of $\theta = 115\pm8^\circ$ for several amounts of secondary fluid. }
\label{fig:histograms_115}
\end{center}
\end{figure*}

The average values for the $\theta = 115\pm8^\circ$ series of samples, i.e. the capillary state system, are shown in Figure~\ref{fig:averages_cap} 
and histograms of coordination number and clustering coefficient are shown in Figure~\ref{fig:histograms_115}.

The average coordination number for the sedimented bed $\bar{z} = 3.3$ is a lower than for the pendular state samples and the clustering coefficient, has the same average value of $\bar{c}=0.15$. The histograms for this first $\theta = 115\pm8^\circ$ point (Figure \ref{fig:histograms_115}b.I) resembles that of the $\theta = 87\pm8^\circ$ sample with wetted asperities ($\phi_{\mathrm{sec}} / \phi_{\mathrm{solid}} = 0.002 \pm 0.001$, Figure~\ref{fig:histograms_87}II)

Interestingly, the development of a network in the capillary state seems to follow a different pattern than in the pendular state. Instead of an increasing amount of particles with clustering coefficient near zero, i.e. binary contacts, one can see that the amount of such particles is reduced in this sample. Instead, there is a large percentage of particles with intermediate or high clustering. The distribution between the clustering coefficient ranges is again shown in Figure S10 of the supplementary material. With the addition of secondary fluid, the average coordination number and average clustering coefficient both rise. This is consistent with both computational models of the capillary state where particles surround small fluid droplets forming polytetrahedrons\cite{koos2014capillary, fortini2012clustering} and is also consistent with the confocal images (Figure~\ref{fig:Images_115}) showing particle clusters.

The histograms reveal that even with small amounts of secondary liquid, the overall distribution of the coordination number is broader compared to the pendular state samples. Furthermore, the peak at $c = 0.3$ is much more pronounced in early stages of the network buildup. Together with the lower amount of particles at $c = 0$, we conclude that much fewer binary contacts are present. Overall, the histograms do not change much with the addition of secondary liquid, although the averages increase slightly, indicating that structures present in the sample are growing while the inner arrangement of the structures themselves remains unaltered and does not undergo significant phase changes.

A special effect can be seen between $\phi_{\mathrm{sec}} / \phi_{\mathrm{solid}} = 0.193 \pm 0.009$ (Figure~\ref{fig:Images_115}.VII) and $\phi_{\mathrm{sec}} / \phi_{\mathrm{solid}} = 0.451 \pm 0.037$ (Figure~\ref{fig:Images_115}.VIII). Between these points, the network transitions from a network of small clusters with many connections to a bicontinuous structure with large clusters and few connections. Between these two points, the average clustering coefficient remains unchanged, but there is a shift from low clustering ($c \leq 0.2$) to medium clustering ($0.2 < c \leq 0.4$) as shown in Figure S3. There is also an increase in the average coordination number from $\bar{z}=5.4$ to $\bar{z}=6.2$, driven by a nearly uniform shift in the histogram to the right. 

As with the pendular state samples, phase separation takes place at high $\phi_{\mathrm{sec}} / \phi_{\mathrm{solid}}$ in the capillary state system. The coordination number reaches a maximum of $\bar{z} = 7.6$ and the clustering coefficient returns to a value of $\bar{c} = 0.26$.  Both the shape of the histograms and the average values indicate that this separated state is essentially the same between the pendular and the capillary state samples.

\subsection{Connecting microstructure and rheology}
The shear moduli are a measure of the momentum transport within the sample. Oscillatory shear experiments allows us to decompose these stresses into a viscous part and into an elastic part. This stresses originate in all systems either by forces acting   or between particles or they are determined by the momentum of the particles themselves. Irving and Kirkwood proposed that the volume averaged stress tensor can be calculated by\cite{irving1950statistical}
\begin{equation}
\sigma_{(k,l)} = \frac{\sum_{i} {m_{i}v_{i}^{(k)}v_{i}^{(l)}}}{V} + \frac{\sum_{j>i}{F_{ij}^{(k)}r_{ij}^{(l)}}}{V}
\end{equation}
where we sum up all individual particle contributions in terms of momentum $v_{i}v_{i}$ and pair forces $F_{ij}$ between particles separated by the distance vector $r_{ij}$. The upper indices $k$ and $l$ describe the position in the stress tensor. As the particles in these samples are non-brownian the structure of the network and the strength of the bonds determine the rheological response of the network together with the hydrodynamic effects.

\subsubsection{Pendular state samples}
\begin{figure}[htbp]
\begin{center}
\includegraphics[width=0.5\textwidth]{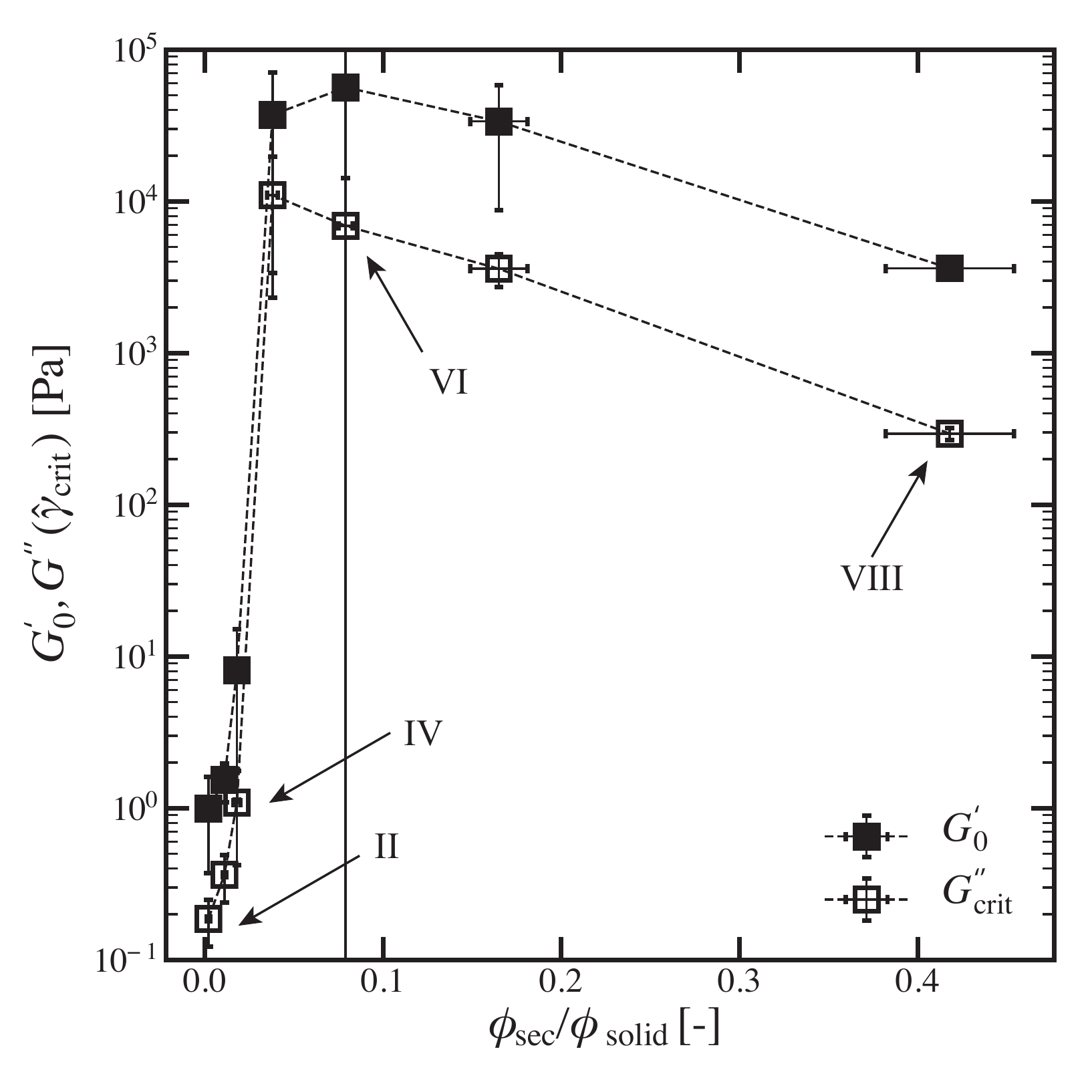}
\caption{Plateau storage modulus $G'_0$ and loss modulus value $G''(\gamma_{\mathrm{crit}})$ at the end of the LVE region, as function of the effective secondary fluid to solid volume ratio $\phi_{\mathrm{sec}} / \phi_{\mathrm{solid}}$ 
for the samples with particles having $\theta = 87\pm8^\circ$.
The frequency was fixed at $\omega = 10$~rad/s.}
\label{fig:modulus_pend}
\end{center}
\end{figure}

\paragraph{Experimental data}
Since capillary suspensions form strong gels with frequency independent moduli, they can be characterised using an oscillatory amplitude sweep. The linear viscoelastic plateau value of the storage modulus, $G'_0$, as well as the loss modulus at the critical strain $G''(\gamma_{\mathrm{crit}})$ of the  $\theta = 87\pm8^\circ$ series of samples are shown in Figure~\ref{fig:modulus_pend} as a function of the ratio of secondary fluid volume fraction $\phi_{\mathrm{sec}}$ to the solid volume fraction $\phi_{\mathrm{solid}}$. The two phase separated samples ($\phi_{\mathrm{sec}} / \phi_{\mathrm{solid}} = 0.0$ and $\phi_{\mathrm{sec}} / \phi_{\mathrm{solid}} = 0.848 \pm 0.076$, points I and IX) are not shown in the rheological data.

A steep increase in the suspensions strength occurs for low amounts of $\phi_{\mathrm{sec}} / \phi_{\mathrm{solid}}$. Both $G'_0$ and $G''(\gamma_{\mathrm{crit}})$ show a maximum at an intermediate amount of added secondary fluid, albeit at slightly different $\phi_{\mathrm{sec}} / \phi_{\mathrm{solid}}$ values. After this maximum, the network strength begins to decrease again. This behaviour is characteristic for pendular state samples.\cite{dittmann2014micro, yang2017preparation, bossler2018fractal} As was previously hypothesised and confirmed with confocal images,\cite{koos2012particle, bossler2016structure} the stress increase at relatively low amounts of secondary fluid should appear due to the buildup of binary capillary bridges, where the maximum is attributed to a coalescence of adjacent bridges, termed the funicular state. The decrease for relatively high amounts of secondary fluid is explained by the formation of large and dense particle agglomerates (e.g. Figure \ref{fig:Images_87}VIII) that begin to destabilise the structure.\cite{bossler2018fractal, dittmann2014micro}

Of particular interest in the pendular state is the transition to the funicular state, which should correspond to the maximum stress values in the rheological data of Fig.~\ref{fig:modulus_pend} in the range of $0.038 \pm 0.003 \leq \phi_{\mathrm{sec}} / \phi_{\mathrm{solid}} \leq 0.079 \pm 0.003$. This value also coincides with a transition from binary bridges to a clustered structure shown in (Figure \ref{fig:Images_87}V and VI). Thus, the decrease in strength observed in the rheological data indeed correlates to the formation of large agglomerates which would reduce the number of network contacts. 
Point VI also corresponds to the end of the rapid rise region in $\bar{z}$ and $\bar{c}$. Points V and VI show a slight broadening of the coordination number histogram and a clear decrease in the number of particles with low or zero clustering and an increase in the particle fraction with intermediate or high clustering. Indeed, this metric shows that the clustering actually begins earlier -- at point IV, $\phi_{\mathrm{sec}} / \phi_{\mathrm{solid}} = 0.018 \pm 0.002$ -- when the fraction of particles with $c=0$ begins to decrease. This may not be manifested in the shear moduli due to the small fraction of clustered particles, but may effect other rheological measures such as yielding.

\paragraph{Elastic stresses}
Elastic stresses can only be caused by the formation of bonds between particles. The amount of bonds is, on the one hand, a function of the volume of added secondary liquid. On the other hand, the arrangements of these bonds is crucial for stress development. Obviously, there is a difference between a droplet of secondary liquid that has been well-dispersed during the sample preparation and the case where only few, large bridges are formed from bad mixing. We propose, therefore, that the elastic stress in our system can be approximated from the average coordination number of particles in the network.\cite{semprebon2016liquid}

Contacts alone, however, are insufficient to create a high elastic modulus. In capillary suspensions with hard spheres, momentum transport contributing to the elastic response has to originate from the secondary liquid. Without secondary liquid, there might be a lot of particle contacts, but these particle contacts rely solely on friction to transmit momentum. Friction forces mostly act perpendicular to the position vector connecting the particles centres via their contact point and, thus, do not contribute to the off-diagonal elements i.e. the shear stress. Adding secondary liquid to fill the particles' asperities makes these contacts \emph{effective} for the transmission of normal forces. Thus, the modulus of the samples begins to increase despite the average coordination number being lower than the one measured in the sedimented bed state. 
Increasing the amount of secondary liquid beyond the point of filled asperities decreases the average coordination number and seems to contradict our hypothesis at first sight. However, the lowering of this average coordination number is mostly caused by the vanishing number of particles with a high contact number and not by a shift in the entire distribution (Figure~\ref{fig:histograms_87}).
Furthermore, samples with a $\bar{z} = 4$ (point V) fulfils the condition of isostaticity for spheres with friction which can explain the steep rise in moduli as well.\cite{shundyak2007force, Hsiao2012role}

The relationship between an increasing number of bonds increasing the transmission of momentum in the sample holds true through the peak $G'_0$ until point VII ($\phi_{\mathrm{sec}} / \phi_{\mathrm{solid}} = 0.165 \pm 0.016$) as shown in the supplementary data Figure S11.
Here, an interesting effect can be observed. First, the average coordination number decreases. However, the average clustering coefficient continues to increase. That means more closed aggregates are formed. These closed aggregates are analogous to the cliques in a small world network. These local clusters redistribute energy inside themselves, as previous works of graph theory have shown.\cite{kuperman2001small} The clusters distribute the motion between their many contacts, rearranging the structure and dissipating any motion. The energy, thus, becomes unavailable. This unavailable energy leads to a drop in the modulus that is not compensated for by the subsequent rise in the coordination number. This is due to the fact that these aggregates become more and more cliquish with high internal interconnectivity, but few contacts bridging the flocs. As already mentioned in the introduction we expect the physical networks to not show an overall clustering coefficient of $\bar{z} = 1$ due to restraints of spatial embeddedness. However, is seems evident that the movement of one particle in such an arrangement can influence the movement of all other particles that are part of the same local cluster. This corresponds to the concept of a clique that connects beyond spatial embeddedness. Similar results have already been observed in optical tweezer experiments. Here, a transition from a cluster-like to a string-like gel was observed indicating a decrease in the long-range structural heterogeneity along with the the occurrence of anomalous strain fields.\cite{lee2008response} At the last point the clustering coefficient decreases with the moduli which has been previously reported to be an indication for material failure, or in the case here, phase separation.\cite{papadopoulos2016evolution}

\paragraph{Viscous stresses}
Viscous stresses, following the strain rate rather than the strain, are caused by skin friction on the particles from the movement of the bulk fluid itself as well as drag and other hydrodynamic effects between the particles and the bulk fluid.

In analogy to the elastic response, we can tie the initial rise in the viscous stresses to the fact that structure formation yields a state where many individual particles are surrounded by bulk fluid increasing the \emph{external} surface area. This increases both the amount of fluid participating in the friction effects on the surface of particles and the amount of primary liquid in the void between the particles, generating stress. This effect seems to cause a rise in the viscous modulus proportional to the rise in the storage modulus for the low amounts of added secondary fluid before point III ($\phi_{\mathrm{sec}} / \phi_{\mathrm{solid}} = 0.011 \pm 0.001$).
However, the loss and storage moduli reach their maxima at point V ($\phi_{\mathrm{sec}} / \phi_{\mathrm{solid}} = 0.038 \pm 0.003$) and point VI ($\phi_{\mathrm{sec}} / \phi_{\mathrm{solid}} = 0.079 \pm 0.003$), respectively. We believe this to be an indication of the pendular-funicular transition that means the merging of binary bridges of secondary liquid to larger interconnected structures. This process decreases the fraction of particles with $c = 0$, leaving only particles of higher clustering with the most prominent peak at $c = 0.3$. The formation of larger and larger clusters continues to decrease the surface area, causing the drop in skin friction and $G''$ in the system. This pendular to funicular transition is marked by a change from a positive correlation between the structure parameters and dynamic moduli to a negative correlation, as shown in Figure S11.

\subsubsection{Capillary state samples}
\begin{figure}[htbp]
\begin{center}
\includegraphics[width=0.5\textwidth]{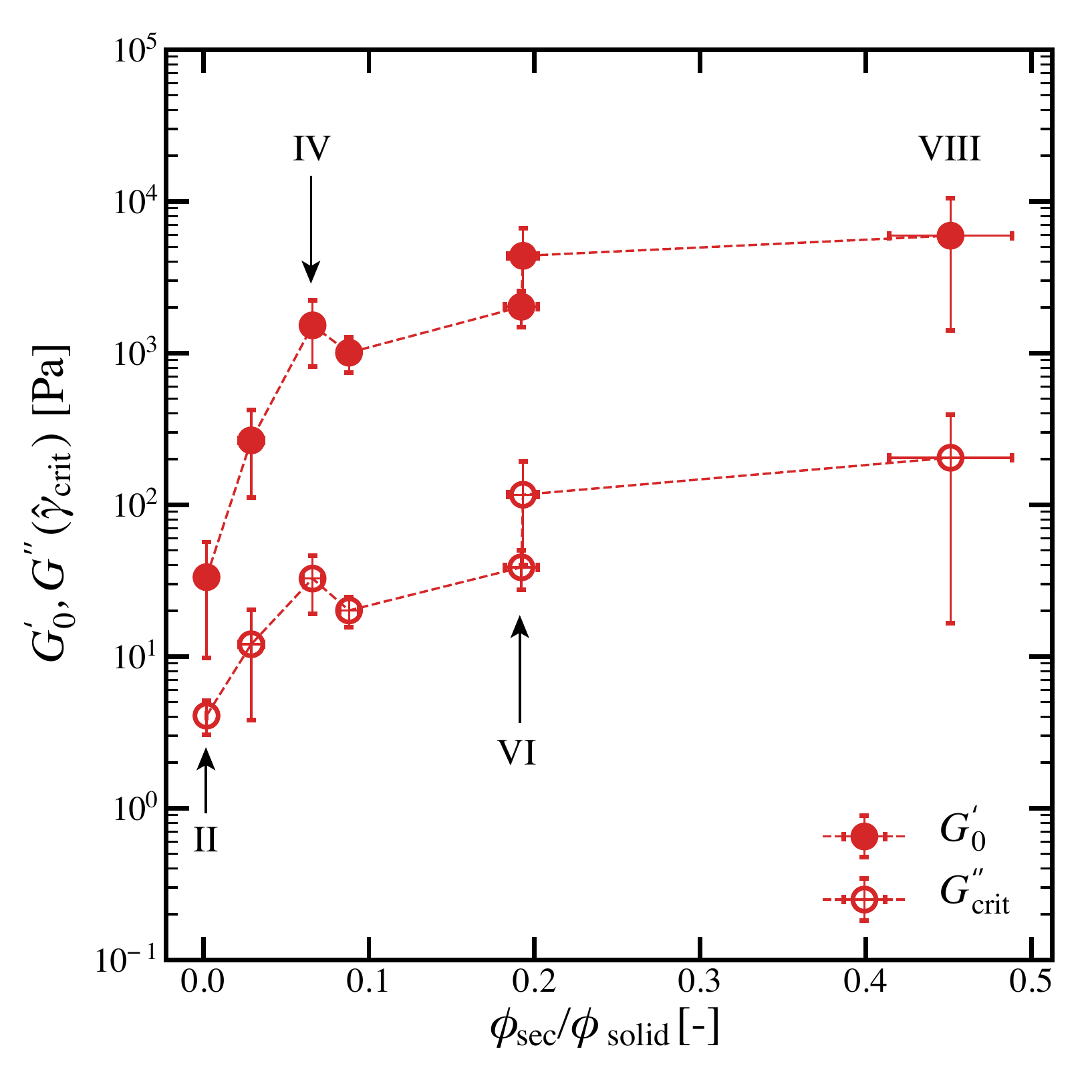}
\caption{Plateau modulus $G'_0$ and loss modulus $G''(\gamma_{\mathrm{crit}})$ at the end of the LVE region, as function of the effective secondary fluid to solid volume ratio $\phi_{\mathrm{sec}} / \phi_{\mathrm{solid}}$ 
for the samples with particles having $\theta = 115\pm8^\circ$.
The frequency was fixed at $\omega = 10$~rad/s.}
\label{fig:modulus_cap}
\end{center}
\end{figure}

The rheological quantities for the $\theta = 115\pm8^\circ$ series are shown in Figure~\ref{fig:modulus_cap}.
Again, the phase separated samples (points I and IX) are not shown.
Unlike for the pendular state sample, the capillary state does not exhibit a  pendular-funicular transition or any other characteristic transitions with the exception of the highest and lowest $\phi_{\mathrm{sec}} / \phi_{\mathrm{solid}}$ (not shown). Instead, a gradual increase of the storage and loss moduli is observed in the rheological data of Fig.\ref{fig:modulus_cap}, where also no maxima or other hints for clear transitions were observed. This finding fits very well the more gradual structural changes, which mostly just show an increasing size of bridges and droplets with increasing secondary fluid volume. 

The connection between the sample microstructure and the resulting rheological properties in the capillary state resides on the same foundation as the pendular state, regardless of the difference in contact angle. Indeed, we can again observe a direct correlation between the main coordination number, the clustering coefficient and the presented moduli. Even the structural transition between $\phi_{\mathrm{sec}} / \phi_{\mathrm{solid}} = 0.088 \pm 0.006$ (point V) and $\phi_{\mathrm{sec}} / \phi_{\mathrm{solid}} = 0.192 \pm 0.006$ (point VI) is well reflected in the rheological as well as the structural data (see Figure S11). As the overall ratio between $\bar{z}$ and $\bar{c}$ does not change with the addition of secondary liquid, we would not expect sudden changes in the moduli. This is indeed the case here. Only upon phase transition at the very end an extreme change in the structure parameters can be observed. Previous comparisons between samples showed that the elastic modulus or yield stress was lower in capillary state samples compared to the pendular state.\cite{bossler2018fractal} We would argue that this difference can arise from the differences in the cliquishness alone, and thus, these measures may not indicate differences in bridge strength. Indeed, in the present samples, the pendular samples with low amounts of $\phi_{\mathrm{sec}} / \phi_{\mathrm{solid}}$ (lower coordination number $\bar{z}$) are weaker than in the capillary state and the bicontinuous samples have nearly identical structural parameters and moduli.

\section{Alternative approaches}
Graph theory tools have recently gained popularity and, thus, have also been applied to other material classes. The force chains of granular matter were investigated\cite{kollmer2019betweenness}. Kollmer and Daniels utilise the measure of betweenness centrality to measure and predict the force response of granular material to uniaxial loading. However, our use of the same measures, as presented in the supplementary information Figure S12 and S13, does not yield conclusive results due to the large error bounds, which are  much higher than for measures such as the clustering coefficient. A reason for this might lie in the differences in the underlying network structures. Grains within granular networks have more uniform spatial distribution compared to the capillary suspensions where large void volumes can occur. Furthermore, capillary suspensions do not experience an ensemble of configurations comparable to the ones present in granular matter. Probing capillary suspension, especially outside of the linear regime, changes their structure irreversibly. This effect does not occur with such dense granular matter as large movement is prohibited by the tight packing of the material.

A further example is the study of depletion interaction gels.\cite{whitaker2019colloidal} In this work, Whitaker et al., studied the structure and rheology of depletion interaction gels with varying attractive potential. The contact number of particles was investigated the same way presented here. However, the mean and width of the coordination number distribution only changed slightly upon an increase of the attractive strength. While not explicitly measured, there is also likely less variation in the clustering coefficient since the internal volume fractions of the cluster only varied from 0.3 to 0.35. Capillary suspensions offer much more versatility as a model system due to the possibility to actively tune the amount of particle neighbours via the amount of secondary liquid added and the mixing conditions. The approach of l-balanced graph partition offers great possibilities for capillary suspensions, e.g., in the regions of of high secondary volume fractions. It comes, however, with the caveat that the amount of particles in the such a decomposition has to be chosen. Especially in the region of linear arrangements, where not a lot of pronounced particles groups are present, this choice remains rather arbitrary. Furthermore, it introduces a further step in the processing, that is not necessary when, e.g., the clustering coefficient is used. Perhaps, however, the combination of the two methods can be used to expand the robustness and insights provided by each method.

\section{Conclusions}
In this paper, we have shown the presence of different structural transitions in capillary suspensions and their impact on the resulting dynamical properties. For this purpose, two sample series containing particles with different contact angles ($\theta = 87 \pm 8^\circ $ and $115 \pm 8^\circ$) were evaluated. Using methods from graph theory, we showed that a distinction between pendular and capillary state might be misleading since the resulting structural properties are, in both cases, very well predicted by local and semi-local measures. Despite the simplicity of this approach, the obtained results are very intriguing in their nature. Both the number of bonds and the cliquishness of clusters, as captured by the coordination and the clustering coefficient, respectively, can be tied tightly to the resulting bulk rheological properties. This method helps, via a handy relationship, to complete the trinity of rational product design, physico-chemical properties and consumer satisfaction which still remains crucial in engineering tasks.
Nevertheless, there is still room for optimisation. First, the results could be improved by taking the strength of the bridges between particles into account. The construction of a weighted graph based on the amount of secondary fluid connecting particles could yield further insights. Other measures offered by graph theory such as centrality have been linked to force networks and could be of interest, but may be harder to quantify for sparse networks. Other larger scale measurements, such as the quantification of the networks backbone using the network's minimum spanning tree, can also be used to characterise the shortest paths in the network transmitting the applied stresses.

\section*{Conflicts of interest}
There are no conflicts to declare.

\section*{Acknowledgements}
The authors would like to thank Joost de Graaf from the University of Utrecht for helpful discussions during the preparation of the manuscript. We would like to acknowledge financial support from the Research Foundation Flanders (FWO) Odysseus Program (grant agreement no. G0H9518N) and from the International Fine Particle Research Institute (IFRPI). 

\bibliographystyle{acm}
\bibliography{SecVol_V2}

\end{document}